\documentclass[%
 reprint,
 amsmath,amssymb,
 aps,
]{revtex4-1}

\usepackage{graphicx}
\usepackage{dcolumn}
\usepackage{bm}

\begin{document}

\preprint{APS/123-QED}

\title{Nonlinear Optics of Optomagnetics;
Quantum and Classical Treatments}

\author{A. Hamed Majedi}
 \altaffiliation[Also at ]{Perimeter Institute for Theoretical Physics, Waterloo, Ontario, Canada.}

 \email{ahmajedi@uwaterloo.ca}
\affiliation{%
Department of Electrical and Computer Engineering and Department of Physics and Astronomy,\\ Waterloo Institute for Nanotechnology, University of Waterloo, Waterloo, Ontario, Canada, N2L 3G1}%

\author{Brahim Lounis}
\affiliation{ Laboratoire Photonique, Numérique et Nanosciences (LP2N), 
Institut d'Optique Graduate School\\ CNRS and Universit\'e de  Bordeaux\\ 
 LP2N- Institut d'Optique d'Aquitaine
Rue Francois Mitterand
F-33400 Talence Cedex, France
}%

\date{\today}

\begin{abstract}
Optomagnetics emerges as a growing field of research cross-linking optics, magnetism and material science. Here, we provide a microscopic quantum mechanical and a macroscopic classical models to describe optomagnetic effects from nonlinear optics point of view. Our self-consistent quantum mechanical formulation considers all orders of perturbing field and results not only in finding generalized Pitaevskii's relationship, where photoinduced magnetization can be expanded in terms of light power, but also provides compact and analytical expressions for optical gyration vector coefficients. classical treatment is then developed based on the anharmonic Drude-Lorentz model showing that the photo-induced DC magnetization is proportional to odd harmonics of the light power. The difference in quantum and classical results are revealed and discussed. Having a pomp-probe setup in mind, we describe how a probe light signal can propagate down an optomagnetic medium, i.e. a medium that is magnetized by intense circularly-polarized pump light, via its  permittivity tensor and find light propagation characteristics. Inverse Faraday and Cotton-Mouton Effects are discussed as a result of circular and linear birefringences and their Verdet constants have been analytically found.   
\begin{description}
\item[Usage]
Secondary publications and information retrieval purposes.
\item[PACS numbers]
May be entered using the \verb+\pacs{#1}+ command.
\end{description}
\end{abstract}

\pacs{Valid PACS appear here}
\maketitle


\section{Introduction}
Nonlinear optics continues to play a central role not only in the advancement of optical sciences and photonic technologies but also to provide powerful tool to probe the structure and properties of materials. Initiated with the series of experiments, direct optical generation and control of magnetization revives the less-explored field of optomagnetics and recreates an exciting synergy between photonic and magnetism communities \cite{kimel2004laser,kimel2005ultrafast,kirilyuk2010ultrafast,subkhangulov2016terahertz,ghamsari2016nonlinear}. The origin of optomagnetics stems from a nonlinear interaction of not-linearly polarized intensive light with the orbital and spin moments of electronic structure where light angular momentum and gyration rules the generation, control, processing and detection of magnetization in matter. Optomagetic effect is mostly understood and explored in light of Inverse Faraday Effect (IFE), i.e. the generation of static magnetization by circularly polarized light.  
IFE has been first predicted by L.P. Pitaevskii in 1960  based on a  phenomenological ansatz on a ground of a generalized Maxwell-Abraham stress tensor in a transparent dispersive medium \cite{pitaevskii1961electric}. He predicted that the static magnetization is related to the optical field intensity through an optical gyration coefficient $\gamma$, in the form of 
\begin{equation}
\label{pita}
{\bf M}_{DC}=\gamma {\bf E}\times{\bf E}^*
\end{equation}
where ${\bf E}$ is the complex electric field intensity. The first experimental observation of the so-called Pitaevskii's relationship has been carried out by J. P. van der Ziel, P. S. Pershan and L. D. Malmstrom and the term Inverse Faraday Effect (IFE) was coined by them in 1965 \cite{van1965optically}. They have provided the quantum mechanical model of IFE based on the effective Hamiltonian method at low frequency limit and justified the Pitaekskii's relationship \cite{pershan1966theoretical}. After a renewed interest in ultrafast optical control of  magnetization \cite{kalashnikova2015ultrafast}, a new theoretical attempts has been initiated for quantum modeling of optomagnetics. The effective Hamiltonian method is further considered in the time domain for the Gaussian-shaped laser pulse to study transient magnetization by D. Popova, A. Bringer and S. Bl\"ugel \cite{popova2011theory}. K. Tagushi and G. Tatara introduced quantum mechanical Green's function formalism to include spin and spin-orbit contributions to the photo-induced magnetization in THz frequencies. Their work explicitly revealed the equal contribution of orbital and spin magnetization in Pitaevskii's relationship \cite{taguchi2011theory,taguchi2012theory}. Based on the perturbative solution of Liouville-von Neumann equation for a generalized nonlinear light-matter interaction, M. Battiato, G. Barbalinardo and P.M. Oppeneer provided an exact solution of photoinduced static magnetization up to the second order in the electric field intensity  \cite{battiato2012beyond,battiato2014quantum}. Their density matrix formulation highlights the various physical effects arising from diagonal and off-diagonal elements due to coherence between different levels and state occupation while the dephasing is phenomenologically considered.\\  
This paper articulates the theory of optomagnetism by focusing on its fundamental physics, finding the generalized Pitaevskii's relationship and the light propagation in optomagnetic media. Both quantum mechanical and classical treatments are presented in detail. We reveal the difference between quantum and classical treatments where classical prediction gives incomplete description from the perspective of generalized Pitaevskii's relationship.  We then focus on how a weak electromagnetic field, i.e. probe signal, is copropagating down a photomegnetic medium, i.e. the medium that is magnetized by intensive circularly-polarized light. Our formulation provides the permittivity tensor for optomagnetic medium and attenuation and propagation constants for probe signal leading to the definition of Verdet constant of IFE and rotatory power of inverse Cotton-Mouton effect.\\
\section{Quantum Mechanical Theory of Optomagnetism}
\label{QM}
The underlying physics of optomagnetism for an atomic system can be captured by solving Schr\"odinger equation under the influence of circularly-polarized electromagnetic field. The optical field ${\bf E}_p(t)$ as a discrete sum of positive and negative frequency components of the pump frequency, i.e.   $\omega_p$, is considered in the form of  
\begin{eqnarray}
\label{Efield}
{\bf E}_p(t)&=&{\bf E}_p(\omega_p)e^{i\omega_p t}+c.c.\\
&=&\frac{1}{2}E_{o}(\omega_p)({\bf x}+i{\bf y})e^{i\omega_p t}+c.c.
\end{eqnarray}
where ${\bf x}, {\bf y}, {\bf z}$ are the Cartesian unit vectors and $E_o(\omega_p)$ is the real amplitude of the electric field. The atomic spinor wavefunction $\Psi({\bf r},t)$ is the solution to the following time-dependent Schr\"odinger equation
\begin{equation}
\begin{aligned}
\label{SE}
i\hbar\frac{\partial}{\partial t}\Psi({\bf r},t)=
\hat{H_o}\Psi({\bf r},t)-\hat{\mu}{\bf E}_p(t)\Psi({\bf r},t)\\
=\hat{H_o}\Psi({\bf r},t)-e\hat{{\bf r}}{\bf E}_p(t) \Psi({\bf r},t)
\end{aligned}
\end{equation}
which is written in terms of the sum of Hamiltonian $\hat{H_o}$ for a free atomic system and dipole interaction Hamiltonian, where $e$ is the electron charge and $\hat{\mu}$ and $\hat{{\bf r}}={\bf x}\hat{x}+{\bf y}\hat{y}+{\bf z}\hat{z}$ are the dipole moment and position vector operators, respectively. We seek the general solution to equation (\ref{SE}) using perturbation theory, namely Rayleigh-Schr\"odinger method, where the atomic spinor wavefunction can be written in the following expansion
\begin{equation}
\label{perturb}
\Psi({\bf r},t)= \sum_{N=0}\lambda^N\Psi^{(N)}({\bf r},t)
\end{equation}
and $\lambda$ is the perturbation parameter set in the Hamiltonian in the form of $\hat{H_o}-\lambda\hat{\mu}{\bf E}_p(t)$ \cite{boyd2019nonlinear}. Assuming that the atomic system rests initially in its nondegenerate ground state, i.e. $\Psi^{(0)}({\bf r},t)$, with the energy $E_g=\hbar\omega_g$ given by 
\begin{equation}
    \Psi^{(0)}({\bf r},t)=u_g({\bf r})e^{i\omega_g t}
\end{equation}
the remaining terms in the perturbation expansion (\ref{perturb}) obey the following expression
\begin{equation}
\label{psi_N}
i\hbar\frac{\partial}{\partial t}\Psi^{(N)}({\bf r},t)=
\hat{H_o}\Psi^{(N)}({\bf r},t)-e\hat{{\bf r}}{\bf E}_p(t)\Psi^{(N-1)}({\bf r},t)
\end{equation}
where $N$ is an integer number. Note that $u_g({\bf r})$ represents the stationary ground stat spinor of the atomic system in the absence of any electromagnetic interaction. The solution to equation (\ref{psi_N}) can be written as the summation of the atomics' eignefunction spinors, i.e. $u_l({\bf r})$, 
\begin{equation}
    \Psi^{(N)}({\bf r},t)=\sum_la^{(N)}_l(t)u_l({\bf r})e^{i\omega_l t}
\end{equation}
with the time-dependent probability amplitude $a^{(N)}(t)$ as
\begin{equation}
\label{prob_amp}
    a^{(N)}_m(t)=\frac{1}{i\hbar}\int\limits_{-\infty}^t \sum_l a^{(N-1)}_l(t')u_l({\bf r})V_{ml}(t')e^{-i\omega_{ml}t}dt'
\end{equation}
where the interaction Hamiltonian for the circularly-polarized light and two-dimensional cross-section of the atomic system is 
\begin{eqnarray}
\label{inter}
\nonumber
    V_{ml}(t)&=&- \frac{e}{2}E_o(\omega_p)e^{i\omega_p t}\langle u_m({\bf r})|\hat{ x}+i\hat{y}|u_l({\bf r})\rangle+c.c.\\
    &=&-\frac{e}{2}E_o(\omega_p)r_{ml}e^{i\omega_p t}+c.c.
\end{eqnarray}
To obtain the DC magnetization based on the perturbed eignefunction spinors, one can use the magnetization operator ${\hat{{\bf M}}}$ as 
\begin{eqnarray}
\label{DCMagnet1}
\langle \hat{{\bf M}}\rangle_{DC}&=&
\sum_N\langle\Psi^{(N)}|\hat{{\bf M}}|\Psi^{(N)}\rangle\\
&=&
\nonumber
\sum_{Nl} \bigg|a^{(N)*}_l(t)a^{(N)}_l(t)\bigg|_{DC}\int u_l^*({\bf r})\hat{{\bf M}}u_l({\bf r})d^3{\bf r}
\end{eqnarray}
where $|a^{(N)*}_l(t)a^{(N)}_l(t)|_{DC}$ is the DC terms in the probability density, i.e. $a^{(N)*}_l(t)a^{(N)}_l(t)$ and 
\begin{equation}
\label{Mag}
\hat{{\bf M}}=\displaystyle{\frac{Ne}{2m}(\hat{{\bf L}}+g_s\hat{{\bf S}})}
\end{equation}
is considered as the summation of the angular momentum, $\hat{{\bf L}}$, and spin, $\hat{{\bf S}}$, operators \cite{band2006light}. 
In expression (\ref{Mag}), $N$ is the electron's number density that is exposed to the light, $m$ is the mass of electron and $g_s$ is the electron spin {\it g}-factor.  
In order to explicitly express the photoinduced DC magnetization in terms of the optical field via the interaction Hamiltonian, i.e. equation (\ref{inter}), equation (\ref{DCMagnet1}) can be written as
\begin{eqnarray}
\label{DCmagnet2}
\nonumber
    \langle \hat{{\bf M}}\rangle_{DC}&=&{\bf m}^{(0)}+
    {\bf m}^{(1)}|E_o(\omega_p)|^2+{\bf m}^{(2)}|E_o(\omega_p)|^2|E_o(\omega_q)|^2\\&+&{\bf m}^{(3)}|E_o(\omega_p)|^2|E_o(\omega_q)|^2|E_o(\omega_r)|^2+...
\end{eqnarray}
where $\omega_p$, $\omega_q$, $\omega_r$, ... are the pump frequencies and ${\bf m}^{(i)}$ are introduced as the $i^{th}$ order optical gyration vectors. This equation explicitly shows that the photo-induced DC magnetization can be expanded as a power series of pumping intensity. The optical gyration vectors crucially depends on the detail of the dipole moment vector, pump frequencies  and the detail of eigenenergy spinors through equation (\ref{DCMagnet1}). Considering the circularly-polarized light, i.e. equation $(\ref{Efield})$, propagating in the $z$ direction,  and the transition dipole moment in $x-y$ plane, i.e. equation (\ref{inter}), the expansion of the magnetization and spin operators in Cartesian coordinate system dictates the direction of optical gyration vectors in $\pm z$ direction.     
${\bf m}^{(0)}$ is related to the collective orbital magnetic moment of the atomic system in its ground state in the absence of any interaction as
\begin{eqnarray}
\nonumber
    {\bf m}^{(0)}&=&\frac{Ne}{2m}\langle u_g|(\hat{{\bf L}}+g_s\hat{{\bf S}})|u_g\rangle\\
   &=&\frac{Ne}{2m}\int u_g^*({\bf r})(\hat{{\bf L}}+g_s\hat{{\bf S}})u_g({\bf r})d^3{\bf r}
\end{eqnarray}
The first order optical gyration vector ${\bf m}^{(1)}$ is the result of the optical field interaction with pump frequency $\omega_p$ that excites the atomic system from its ground state energy $E_g=\hbar\omega_g$ to the $m^{th}$ eigenenergy state $E_m=\hbar\omega_{mg}=\hbar(\omega_m-\omega_g)$ with the probability amplitude $a^{(1)}_m(t)$ and can be written as
\begin{eqnarray}
\label{m1}
\nonumber
{\bf m}^{(1)}&=&
\frac{Ne}{2m}\big(\frac{e}{2\hbar}\big)^2\sum_m\langle u_m|(\hat{{\bf L}}+g_s\hat{{\bf S}})|u_m\rangle|r_{mg}|^2\\
&&\big[D^{-1}(\omega_p-\omega_{mg})+D^{-1}(\omega_p+\omega_{mg})\big]
\end{eqnarray}
where $D$ is the defined as a quantum mechanical dispersion relation for positive and negative pumping frequency, $\omega$, as
\begin{equation}
D(\omega\pm\omega_{ij})\triangleq \big|(\omega\pm\omega_{ij})(\omega\pm\omega^*_{ij})
\big|=(\omega\pm\omega_{ij})^2+\Gamma^2_{ij}
\end{equation}
and $\omega_{ij}$ is crudely updated to incorporate the damping phenomena for transition probability between two energy bands as 
$\omega_{ij}=\omega_i-\omega_j-i\Gamma_{ij}$. $\Gamma_{ij}$ is related to the population decay rate of the upper level $i$ and does not represent the dephasing process or the cascaded population among the excited states.
Note that ${\bf m}^{(1)}$ consists the second harmonic generation magnetization oscillating at $2\omega_p$ as well. 
The second order correction to the probability amplitude can yield the second order gyration vector ${\bf m}^{(2)}$ under the influence of the nondegenerate pumping frequency $\omega_q$ as
\begin{eqnarray}
\nonumber
{\bf m}^{(2)}&=&
\frac{Ne}{2m}\big(\frac{e}{2\hbar}\big)^4\sum_n 
\langle u_n|(\hat{{\bf L}}+g_s\hat{{\bf S}})|u_n\rangle|r_{mg}r_{nm}|^2\\
\nonumber
&&\bigg[D^{-1}(\omega_p-\omega_{mg})D^{-1}(\omega_p+\omega_q-\omega_{ng})\\
\nonumber
&+&
D^{-1}(\omega_p-\omega_{mg})D^{-1}(\omega_p-\omega_q-\omega_{ng})\\
\nonumber
&+&
D^{-1}(\omega_p+\omega_{mg})D^{-1}(\omega_p+\omega_q+\omega_{ng})\\
&+&D^{-1}(\omega_p+\omega_{mg})D^{-1}(\omega_p-\omega_q+\omega_{ng})\bigg]
\end{eqnarray}
 The third order optical gyration vector is given in \ref{app1}. This procedure can be systematically applied to the $N^{th}$ order optical gyration vector. The higher order terms in the photoinduced DC magnetization, i.e. $\langle\hat{{\bf M}}\rangle_{DC}$, are proportional to $(\frac{e|E_o|}{2\hbar})^{2N}$. Although the optical gyration vectors depend on the detail of materials' atomic spectra and band structures but they do not possess any symmetry restrictions. The photo-induced magnetization can be thus allowed in any materials regardless of their electrical, magnetic and optical properties. Knowing the fact that the intensity of circularly-polarized light, expressed by the equation (\ref{Efield}), is proportional to its helicity, i.e. 
${\bf z}|E_o(\omega_p)|^2=2i{\bf E}_p\times{\bf E}_p^*$, equation (\ref{DCmagnet2}) represents the quantum mechanical version of a generalized Pitaevskii's relationship and in the case of degenerate pumping field can be written as:
\begin{eqnarray}
\label{M_DC}
\nonumber
    \langle\hat{{\bf M}}\rangle_{DC}&=&\sum_{N=0}{\bf m}^{(N)}\big|(2i{\bf E}_p\times{\bf E}_p^*)\big|^{N}\\
    \nonumber
    &=&{\bf m}^{(0)}+{\bf m}^{(1)}\big|2i{\bf E}_p\times{\bf E}_p^*\big|+
    {\bf m}^{(2)}\big|2i{\bf E}_p\times{\bf E}_p^*\big|^{2}\\
    &+&
    {\bf m}^{(3)}\big|2i{\bf E}_p\times{\bf E}_p^*\big|^{3}+...
\end{eqnarray}
The generalized Pitaevskii relationship in this case consists of the ground state magnetization in the absence of any electromagnetic radiation and the second term represents the Pitaevskii's relation, i.e. equation (\ref{pita}). This equation predicts that optomagnetic effect should be more pronounced in the material that do not possess ground-state magnetization. \\
 Note that the photoinduced DC magnetization crucially depends on the interaction of not-linearly-polarized light through dipole interaction $r_{ml}$, i.e. equation (\ref{inter}) in case of circular polarization, with the expectation value of the magnetization operator acting on spinor eigenfunctions, $\langle u_l|(\hat{{\bf L}}+g_s\hat{{\bf S}})|u_l\rangle$. Obviously, a linearly polarized electric field does not induce any magnetization as it is evidenced by the interaction Hamiltonian, i.e. equation (\ref{inter}).    
The photo-induced static magnetization presented in equation (\ref{M_DC}) can be also generalized to consider time-varying cases where the magnetization is expressed in terms of pump frequencies $\omega_p, \omega_q,...$ and their harmonics. This can be done by finding the expectation value of the magnetization operators between various eignefunction spinors where their energy differences corresponds to the harmonics of light frequencies. Similar to application of nonlinear optical susceptibility, photomagnetic effect can be employed not only for optical processes such as harmonic generation, up/down conversion, switching and mixing but also for probing magnetic properties of materials.\\
\section{Classical Theory of Optomagnetism}
The first attempt to classically treat the IFE dates back to 1975 when B.A. Zon and V. Ya. Kupershmidt used the Drude-Lorentz model to justify the Pitaevskii relationship \cite{zon1976inverse}. This method is further considered for the free-electron gas by R. Hertel \cite{hertel2006theory,hertel2015macroscopic} and reused to find the Verdet constant associated with IFE by M. Battiato {\it et. al.} \cite{battiato2014quantum}. Hereby, we use the nonlinear Drude-Lorentz model based on anharmonic oscillator model to go beyond Pitaevskii relationship and make a comparison with our quantum mechanical treatment presented in section \ref{QM}.\\
To model the photomagnetic effect, the Drude-Lorentz model is adopted in a nonlinear regime under the influence of high-intensity circularly-polarized light. The local electric field will cause the average position of an electron distribution, i.e. ${\bf{r}}(t)$, to be displaced from its equilibrium. For high-intensity light, a large deviation from the average position is expected and the electrons experience anharmonic potential in the form of 
$\displaystyle{U({\bf r})=\frac{1}{2}m\omega_o^2{\bf r}^2+
\frac{1}{3}ma{\bf r}^3+\frac{1}{4}mb{\bf r}^4}$, where $m$ is the mass of electron, $\omega_o$ is the resonant frequency of the oscillator corresponding to the main observed atomic spectral line, and $a$, $b$ characterize the strength of the anharmonicity \cite{bloembergen1996encounters,boyd2019nonlinear}. For materials exhibit centrosymmetric and non-centrosymmetric inversion symmetry, $a=0$ and $b=0$, respectively \cite{band2006light}.
The equation of motion of the electron position can be taken the form 
\begin{equation}
\label{motion}
m\ddot{\bf r}(t)+m\Gamma\dot{\bf r}(t)+m\omega_o^2{\bf r}(t)+ma{\bf r}^2(t)+mb{\bf r}^3(t)={\bf F}(t)    
\end{equation}
where $e$ is the electron charge, $\Gamma$ is the friction term representing the energy loss associated with the material absorption process, ${\bf F}(t)=e{\bf E}(t)$ is acting force and ${\bf E}(t)$ is the vector electric field associated with light pump in the form of
\begin{equation}
\label{Efield2}
{\bf E}(t)=\mbox{Re}\bigg\{{\bf E}e^{i\omega_p t}\bigg\}=
\mbox{Re}\bigg\{E_o({\bf x}+i{\bf y})e^{i\omega_p t}\bigg\}
\end{equation}
where $Re\{.\}$ denotes the real part of a complex function.
The intensity of the circularly-polarized light creates a helicity of the wave in the plane perpendicular to its direction of propagation given by 
\begin{equation}
\label{helicit}
{\bf z}|E_o|^2=\frac{1}{2}i{\bf E}\times{\bf E}^*   
\end{equation}
that is enforcing a gyrating motion on the electrons.This light-induced localized current density in the region compared to the wavelength of light leads to a magnetic moment density or magnetization as \cite{jackson1999classical}
\begin{equation}
\label{magnet}
{\bf M}(t)=\frac{Ne}{2m}{\bf L}(t)=N\frac{e}{2}{\bf r}(t)\times{\bf v}(t)  
\end{equation}
where $N$ is the number density of electrons exposed to light, ${\bf L}$ is the angular momentum and ${\bf v}(t)=\dot{\bf r}$ is the average electron velocity. Note that the intrinsic angular momentum of electrons that is proportional to their spin cannot be considered in such a classical treatment.\\
To find the magnetization, one needs to solve the nonlinear equation (\ref{motion}) under the influence of the electric field based on the perturbation method analogous to that of presented in section \ref{QM}. 
Using expression (\ref{Efield2}), the equation of motion  (\ref{motion}) has the solution in the form of  
\begin{equation}
\label{solution}
{\bf r}(t)=\mbox{Re}\Big\{\sum_{n=1}^n {\bf r}e^{in\omega_p t}\Big\}=\mbox{Re}\Big\{\sum_{n=1}^n \zeta^n {\bf r}_e^{(n)}E_o^ne^{in\omega_p t}\Big\}
\end{equation}
 where $\zeta$ is the perturbation parameter, ${\bf r}_e^{(n)}={\bf x}x_e^{(n)}+{\bf y}y_e^{(n)}$ is the $n^{th}$ order solution in the frequency domain and $y_e^{(n)}=i^n x_e^{(n)}$ due to circular polarization of the incident light. The magnetization in equation (\ref{magnet}) can be written in the frequency domain 
 \begin{eqnarray}
\nonumber
{\bf M}&=&{\bf z}\frac{Ne\omega}{4}
\mbox{Re}\bigg\{\displaystyle{\sum_{n=1}^n} x_e^{(n)}E_o^ne^{in\omega_p t}.\bigg[
\sum_{n=1}^n ni^{n+1}x_e^{(n)}E_o^ne^{in\omega_p t}\\
\nonumber
&+&
\sum_{n=1}^n n(-1)^{n+1}i^{n+1} x_e^{*^{(n)}}E_o^{*}{^n}e^{-in\omega_p t}\bigg]\\
\nonumber
&-&\displaystyle{\sum_{n=1}^n} ni x_e^{(n)}E_o^ne^{in\omega_p t}.\bigg[
\sum_{n=1}^n i^nx_e^{(n)}E_o^ne^{in\omega_p t}\\
&+&\sum_{n=1}^n (-1)^ni^nx_e^{*^{(n)}}E_o^{*^{n}}e^{-in\omega_p t}\bigg]\bigg\}
\label{magnet1}
 \end{eqnarray}
Equation (\ref{magnet1}) evidently shows that the photomagnetic effect is purely nonlinear phenomenon with respect to the electric field intensity as the DC term is proportional to the light intensity $E_o^2$ and its odd harmonics. \\ 
Using equation (\ref{helicit}), the DC component of the magnetization can be generally written as
\begin{eqnarray}
\label{DCmagnet}
\nonumber
{\bf M}_{DC}&=&\sum_{k=0}^n \gamma^{(2k+1)}\big(\frac{i}{2}{\bf E}\times{\bf E}^*\big)
\big|\frac{i}{2}({\bf E}\times{\bf E}^*\big)\big|^{2k}\\
\nonumber
&=&\gamma^{(1)}\big(\frac{i}{2}{\bf E}\times{\bf E}^*\big)+\gamma^{(3)}\big(\frac{i}{2}{\bf E}\times{\bf E}^*\big)\big|(\frac{i}{2}{\bf E}\times{\bf E}^*\big)\big|^{2}+...\\
\end{eqnarray}
where coefficients, $\gamma^{(2k+1)}$, take the following form
\begin{equation}
\label{gamma}
\gamma^{(2k+1)}=-\omega_p(2k+1) \mbox{Re}\bigg\{i^{2k}|x_e^{(2k+1)}|^2\bigg\}
\end{equation}
and represent the optical gyration coefficients, similar to magnetogyration coefficients \cite{landau2013electrodynamics,saleh2019fundamentals}. Equation (\ref{DCmagnet}) clearly shows that the DC magnetization depends on the odd power of the light intensity or helicity vector irrespective of any symmetry in the structure of the material, a prediction that is an incomplete based on quantum mechanical treatment. It is interesting to compare the optical gyration coefficients with the linear and nonlinear susceptibilities, i.e. $\chi^{(n)}$, based on the power series expansion of the electrical field for polarization, i.e. 
$P(t)=\epsilon_o\big(\chi^{(1)}E(t)+\chi^{(2)} E^2(t)+\chi^{(3)}E^3(t)+...\big)$. We noted that the optical gyration coefficients can be expressed based on susceptibilities as follows 
\begin{equation}
\label{magnet2}
|\gamma^{(2k+1)}|=\frac{(2k+1)\omega_p}{Ne}\Big(\frac{\epsilon_o}{2^{k+1}}\Big)^2|\chi^{(2k+1)}|^2
\end{equation}
Similar prediction, as the third-order nonlinearity induced by IFE  has been made in the context of magnetoplasmonic structures \cite{im2017third}.
This classical prediction is also seen from our quantum mechanical treatment where the optical gyration coefficients are proportional to the square of electrical dipole transition moment, i.e. equation (\ref{m1}). Equation (\ref{magnet2}) shows that material with large linear and nonlinear susceptibilities should exhibit large optical gyration coefficients while they scale linearly with pump optical frequency and inversely with density number of electrons in material. Our model then predicts the optical gyration coefficients in two-dimensional materials should be larger than their bulk counterparts. \\
To fully consider the effect of anharmonicity of the electron's potential, i.e. nonlinear parameters $a$ and $b$, the first three orders of the solution  (\ref{solution}) can be expressed as 
\begin{eqnarray}
y_e^{(1)}&=&ix_e^{(1)}=i\frac{e}{m}\frac{1}{D(\omega_p)}\\
y_e^{(2)}&=&-x_e^{(2)}=a\big(\frac{e}{m}\big)^2\frac{1}{D^2(\omega_p)D(2\omega_p)}
\end{eqnarray}
and
\begin{equation}
\begin{aligned}
y_e^{(3)}=-ix_e^{(3)}=-i\big(\frac{e}{m}\big)^3\bigg(\frac{2a^2}{D(2\omega_p)}-b\bigg)
\frac{1}{D^3(\omega_p)D(3\omega_p)}
\end{aligned}
\end{equation}
where 
\begin{equation}
D(\omega_p)\triangleq\omega_o^2-\omega_p^2+i\omega_p\Gamma
\end{equation}
is the dispersion function of a damped harmonic oscillator. The first two orders of the optical gyration coefficients are then
\begin{equation}
\label{gamma1}
\gamma^{(1)}=-(\frac{e}{2m})\epsilon_o\omega_{pl}^2\frac{\omega_p}{|D(\omega_p)|^2}
\end{equation}
\begin{equation}
\begin{aligned}
\gamma^{(3)}=(\frac{e}{2m})\epsilon_o\omega_{pl}^2\big(\frac{e}{m}\big)^4\frac{3\omega_p}{|D^3(\omega_p)|^2|D(3\omega_p)|^2}\\
\bigg|\big(\frac{2a^2}{D(2\omega_p)}-b\big)\bigg|^2
\end{aligned}
\end{equation}
where $\displaystyle{\omega_{pl}\triangleq\sqrt{\frac{Ne^2}{m\epsilon_o}}}$ is the plasma frequency of the material.
Equation (\ref{gamma1}) is independent of the nonlinear coefficients $a$ and $b$ and is in agreement with the results in \cite{hertel2006theory,battiato2014quantum}. The classical model shows that each optical gyration coefficients is a collective response of $N$ orbiting electrons represented by the plasma frequency $\omega_{pl}$ and the gyromagnetic ratio of electron, i.e. $\displaystyle{\frac{e}{2m}}$ that is dictated by the helicity of the pumping light and is modified by the frequency response of a classical atom as an anharmonic oscillator to the odd harmonics of pumping light intensity. This is the consequence of angular momentum conservation between light and $N$ noninteracting electrons in the presence of anharmonic oscillator representing atomic structure. Evidently, this equation, i.e. equation (\ref{DCmagnet}) is partially inconsistent with the the quantum theory where the photo-induced magnetization is proportional to all harmonics of light intensity, i.e. equation (\ref{M_DC}). This fact stems from the fundamental difference on how the state of electron is considered quantum mechanically by the wavefunctions and operators acting on it through momentum and angular momentum, and classically, by its position vector and its temporal derivative.\\  
It is straight forward to justify that the magnetic field associated with the circularly polarized light, i.e.${\bf B}={\bf z}B_o(-i{\bf x}+{\bf y})e^{i\omega_p t}$, has no contribution to the DC magnetization in the context of our quantum mechanical and classical treatments.
\section*{Light Propagation in Optomagnetic Media}
We are considering the propagation of a weak linearly-polarized optical signal with a frequency of $\omega_s$ in an optomagnetic material where a DC magnetization is induced by the co-propagation of strong circularly-polarized pump light in $z$ direction. The photo-induced magnetic field by the pump light, $B_o=\mu_o M_{DC}$ where $M_{DC}$ is governed by equation (\ref{DCmagnet}), breaks the directional symmetry of the linear dielectric constant for the optical signal similar to magneto-optic material leading to linear birefringence or Cotton-Mouton effect and circular birefringence or Faraday effect. The dependence of the imaginary part of the permittivity, i.e. $\epsilon''_{ij}$, on $B_o$ leads to circular birefringence or Faraday effect, while dependence of the real part of the permittivity, i.e. $\epsilon'_{ij}$, on $B_o$ leads to linear birefringence or Cotton-Mouton effect \cite{shen1984principles}. 
These effect can be described by exploiting Drude-Lorentz model for weak optical signal where the Lorentz force acting on the bound electrons is due to electric field of light signal and DC magnetic field produced by Circularly-polarized light. The solution of linear version of the equation (\ref{motion}), i.e. $a=b=0$, where $F(t)=e({\bf x}E_x+{\bf y}E_y)+ev(t)\times {\bf z}B_o$, leads to anisotropic relative dielectric constant that it can be cast into the following tensor form  
\begin{equation}
\begin{aligned}
\label{dielectric1}
\bar{\bar{\epsilon_r}}(\omega_s)=\left(\begin{array}{ccc} \epsilon'_{xx}-i\epsilon''_{xx} & \epsilon'_{xy}-i\epsilon''_{xy} & 0\\ 
\epsilon'_{yx}-i\epsilon''_{yx} & \epsilon'_{yy}-i\epsilon''_{yy} & 0\\
0 & 0 & \epsilon'_{zz}-i\epsilon''_{zz} \end{array}\right)=\\
\left(\begin{array}{ccc} 1+\displaystyle{\frac{\omega^2_{pl}}{D^2_F(\omega_s)}}D(\omega_s) & -i\omega^2_{pl}\omega_c\displaystyle{\frac{\omega_s}{D^2_F(\omega_s)}} & 0\\ 
i\omega^2_{pl}\omega_c\displaystyle{\frac{\omega_s}{D^2_F(\omega_s)}}& 1+\displaystyle{\frac{\omega^2_{pl}}{D^2_F(\omega_s)}}D(\omega_s) & 0\\
0 & 0 & 1+\displaystyle{\frac{\omega^2_{pl}}{D(\omega_s)}} \end{array}\right)
\end{aligned}
\end{equation}
where 
\begin{equation}
D^2_F(\omega_s)\triangleq\omega_o^2-(1+\omega^2_c)\omega_s^2+i\omega_s\Gamma=D^2(\omega_s)-\omega^2_c\omega_s^2
\end{equation}
is the modified dispersion function due to the presence of magnetic field and $\omega_c=\displaystyle{\frac{eB_o}{m}}=\frac{\mu_o e}{m}M_{DC}$ is the cyclotron frequency that is induced by the circularly-polarized pump light. Note that the $z$-axis is not affected by the photo-induced magnetic field. By expanding the complex permittivity elements in series with respect to the photo-induced magnetic field, $B_o$, we obtain the real and complex parts of the permittivity elements 
\begin{eqnarray}
\label{dielectric2}
\epsilon'_r=\epsilon'_{xx}=\epsilon'_{yy}&\approx&1+\frac{\omega^2_{pl}}{|D(\omega_s)|^2}(\omega^2_o-\omega_s^2)\\
\epsilon''_{xx}=\epsilon''_{yy}&\approx&\frac{\omega^2_{pl}\Gamma}{|D(\omega_s)|^2}\omega_s\\
\epsilon'_{xy}=-\epsilon'_{yx}&\approx& -2\omega^2_{pl}\Gamma\frac{\omega_s^2(\omega^2_o-\omega_s^2)}{|D(\omega_s)|^4}\omega_c\\
\epsilon''_{xy}=-\epsilon''_{yx}&\approx&\omega^2_{pl}\frac{(\omega^2_o-\omega_s^2)^2-\omega_s^2\Gamma^2}{|D(\omega_s)|^4}
\omega_s\omega_c\\
\epsilon'_{zz}&=&1+\frac{\omega^2_{pl}}{|D(\omega_s)|^2}(\omega^2_o-\omega_s^2)\\
\epsilon''_{zz}&=&\frac{\omega^2_{pl}\Gamma}{|D(\omega_s)|^2}\omega_s
\end{eqnarray}
It is worth noting that the Onsager symmetry of the permittivity, i.e. $\epsilon'_{ij}(\omega_s, B_o)=\epsilon'_{ji}(\omega_s,-B_o)$ and $\epsilon''_{ij}(\omega, B_o)=-\epsilon'_{ji}(\omega,-B_o)$, holds but Hermicity of the dielectric constant, i.e. $\epsilon_{ij}(\omega_s, B_o)=\epsilon^*_{ji}(\omega_s,B_o)$, is valid where the damping factor, $\Gamma$, or absorption is absent. Any media described by anisotropic permittivity tensor, i.e. equation (\ref{dielectric1}), has two normal propagation modes with relative permittivities $(\epsilon'_{xx}\pm\epsilon''_{xy})- i(\epsilon''_{xx}\mp \epsilon'_{xy})$. The permittivity tensor is then diagonalized in the coordinate systems with orthogonal unit vectors ${\bf e}_{\pm}=\displaystyle{\frac{1}{\sqrt{2}}({\bf x}\pm i{\bf y})}$ and ${\bf z}$ represented by uniaxial optical symmetry.
The complex propagation constants, $\alpha_{1,2}+i\beta_{1,2}$, for an optical signal are 
\begin{equation}
\alpha_{1,2}+i\beta_{1,2}\approx \frac{k_o}{2\sqrt{\epsilon'_r}}(\epsilon''_{xx}\pm\epsilon'_{xy})\big(1\pm\frac{\epsilon''_{xy}}{2\epsilon'_r}\big) +i
k_o\sqrt{\epsilon'_r}\big(1\mp \frac{\epsilon''_{xy}}{2\epsilon'_{r}}\big)
\end{equation}
where $k_o=\displaystyle{\frac{2\pi}{\lambda_o}}$ is free space wavenumber in terms of wavelength $\lambda_o$. The ability of the optomagnetic medium to rotate the linear polarization of an optical signal that leads to circular birefringence is commonly referred to as inverse Faraday effect and can be found by its rotatory power, $\rho_{xy}$, similar to the magnetooptic media, that is defined by the rotation angle per unit length as
\begin{equation}
\label{rotatory}
    \rho_{xy}=\frac{\beta_1-\beta_2}{2}\approx -\frac{\pi}{\lambda_o}\frac{\epsilon''_{xy}}{\sqrt{\epsilon'_r}}
\end{equation}
The rotatory power is a linear function of the photo-induced magnetic field, $B_o$. The rotatory power of the optomagnetic medium can be defined based on the intensity of the pump light and in the first order can be expressed as function of light pump light intensity $E^2_o$ through the Verdet constant of the inverse Faraday effect, $V_{IFE}$ as:
\begin{equation}
\begin{aligned}
    V_{IFE}(\omega_p,\omega_s)\triangleq \frac{\rho_{xy}}{E^2_o}=-\frac{\pi}{2\lambda_o}\frac{e^2}{m^2}\frac{\omega^4_{pl}\omega_s}{c^2}\frac{\omega_p}{|D(\omega_p)|^2}\\
    \frac{(\omega^2_o-\omega_s^2)^2-\Gamma^2\omega_s^2}{|D(\omega_s)|^4}\frac{1}{\sqrt{1+\frac{\omega^2_{pl}(\omega^2_o-\omega_s^2)}{|D(\omega_s)|^2}}}
\end{aligned}
\end{equation}
Note that the Verdet constant of the inverse Faraday effect is defined as a real quantity that is related to the polarization rotatory power as a function of both light pump and light signal frequencies and it is different than Verdet constant of Faraday effect.\\
The change in the optical refractive index due to presence of static magnetic field  is originally discovered by W. Voigt in 1902 in gases \cite{voigt1902vii} and by A. Cotton and H. Mouton in 1907 for liquids \cite{cotton1907new}. 
Investigation on equations (\ref{dielectric1}) and (\ref{dielectric2}) reveals that the real part of the diagonal  permittivity elements is also altered by the photoinduced magnetic field leading to linear birefringence in both $x-z$ and $y-z$ planes that can be called inverse Voigt or Cotton-Mouton effect. The difference in real part of the permittivity scales with $M_{DC}^2$, as
\begin{equation}
\begin{aligned}
\epsilon'_{xx}-\epsilon'_{zz}\approx \mu^2_o\frac{e^2}{m^2}\frac{\omega^2_{pl}\omega_s^2(\omega_o^2-\omega_s^2)}{|D(\omega_s)|^4}|{\bf M}_{Dc}|^2\\
=\frac{e^4}{4m^4}\frac{\omega^6_{pl}}{c^4}\frac{\omega_s^2(\omega_o^2-\omega_s^2)}{|D(\omega_s)|^4}\frac{\omega^2_p}{|D(\omega_p)|^4}E^4_o
\end{aligned}
\end{equation}
The rotatory power of the linear birefringence can be then calculated as 
\begin{equation}
    \rho_{xz}=-\rho_{yz}=\frac{\rho_{xy}}{2}=-\frac{\pi}{2\lambda_o}\frac{\epsilon''_{xy}}{\sqrt{\epsilon'_r}}
\end{equation}
The inverse Cotton-Mouton effect is weaker than the inverse Faraday effect and its rotatory power is half of the one produced by the inverse Faraday effect.\\
Both inverse Faraday and Cotton-Mouton effects can be used in free-space and integrated photonic systems for all-optical signal processing and nonreciprocal polarization devices without incorporating magnetic devices and characterization setup.  
\section{Conclusions}
Optomagnetics have developed into an expanding research area with a potential of new discoveries in ultrafast magnetism and optics, novel applications in high-speed magnetic recording, information processing and spintronics, as well as probing quantum and 2D materials.\\
We have provided a unified and generalized theoretical frameworks for optomagnetics, both quantum and classical treatments, through a prism of IFE effect.
First, we start quantum mechanical treatment to not only obtain a clear relationship between the photo-induced magnetization and transition dipole moments but also a generalized Pitaevskii's relationship. Using a perturbative method to solve Schr\"odinger equation in the presence of a  circularly-polarized wave,  our method explicitly and compactly finds the optical gyration vectors due to both orbital and spin magnetic moments. The effect of damping phenomena is incorporated in to the theory by introducing excited state's population decay rate. Our formulation can be easily employed for quantum confined structures, i.e. quantum wells, wires and dots.\\ Secondly, We employ the anharmonic Drude-Lorentz model to find the generalized Pitaevskii's relationship and its associated optical gyration coefficients. Comparison between quantum and classical treatments reveals the incompleteness of the classical treatment while it can lay down the basics for description of optomagnetic medium.\\
Lastly, the propagation of linearly-polarized light signal through an optomagnetic medium, that is described by its first order gyration coefficient, is analyzed through a typical pump-probe setup. Our formalism explicitly provides the analytical expressions of Verdet's constants for IFE and inverse Cotton-Mouton effect through the permittivity tensor of optomagnetic medium.\\      
\section{Acknowledgements}
AHM acknowledges financial and hospitality support of IdEx Visiting fellowship program and Laboratoire de Photonique, Num\'erique et Nanosciences (LP2N) at Universit\'e de Bordeaux, Talence Cedex, France. B.L. acknowledges the Institut Universitaire de France.\\ 
\section{Appendix}
\label{app1}
Here, we report the first three-order corrections to the probability amplitude, i.e. equation(\ref{prob_amp}),  under the influence of circularly-polarized light with pump frequencies, $\omega_p, \omega_q, \omega_r$ and interaction Hamiltonian, i.e. equation(\ref{inter}). They read as \begin{equation}
a_m^{(1)}(t)=\frac{e}{2\hbar}E_o(\omega_p)
\bigg[\frac{r_{mg}e^{i(\omega_p-\omega_{mg})t}}{\omega_p-\omega_{mg}}-\frac{r^*_{mg}e^{-i(\omega_p+\omega_{mg})t}}{\omega_p+\omega_{mg}}\bigg]
\end{equation}
\begin{widetext}
\begin{eqnarray}
\nonumber
a_n^{(2)}(t)&=&\big(\frac{e}{2\hbar}\big)^2E_o(\omega_p)E_o(\omega_q)\sum_m \bigg[\frac{r_{mg}r_{nm}e^{i(\omega_p+\omega_q-\omega_{ng})t}}{(\omega_p-\omega_{mg})(\omega_p+\omega_q-\omega_{ng})}
-\frac{r_{mg}r^*_{nm}e^{i(\omega_p-\omega_q-\omega_{ng})t}}{(\omega_p-\omega_{mg})(\omega_p-\omega_q-\omega_{ng})}\\
&-&\frac{r^*_{mg}r_{nm}e^{-i(\omega_p+\omega_q+\omega_{ng})t}}{(\omega_p+\omega_{mg})(\omega_p+\omega_q+\omega_{ng})}
+\frac{r^*_{mg}r^*_{nm}e^{-i(\omega_p-\omega_q+\omega_{ng})t}}{(\omega_p+\omega_{mg})(\omega_p-\omega_q-\omega_{ng})}\bigg]
\end{eqnarray}
\end{widetext}

\begin{widetext}
\begin{eqnarray}
\nonumber
a_{\nu}^{(3)}(t)&=&\big(\frac{e}{2\hbar}\big)^3E_o(\omega_p)E_o(\omega_q)E_o(\omega_r)
\sum_{mn}
\bigg[\frac{-r_{mg}r_{nm}r_{\nu n}e^{i(\omega_p+\omega_q+\omega_r-\omega_{\nu g})t}}{(\omega_p-\omega_{mg})(\omega_p+\omega_q-\omega_{ng})((\omega_p+\omega_q+\omega_r-\omega_{\nu g}))}\\
\nonumber
&+&
\frac{r_{mg}r^*_{nm}r_{\nu n}e^{i(\omega_p-\omega_q+\omega_r-\omega_{\nu g})t}}{(\omega_p-\omega_{mg})(\omega_p-\omega_q-\omega_{ng})((\omega_p-\omega_q+\omega_r-\omega_{\nu g}))}
+\frac{r^*_{mg}r_{nm}r_{\nu n}e^{-i(\omega_p+\omega_q-\omega_r+\omega_{\nu g})t}}{(\omega_p+\omega_{mg})(\omega_p+\omega_q+\omega_{ng})((\omega_p+\omega_q-\omega_r+\omega_{\nu g}))}\\
\nonumber
&-&\frac{r^*_{mg}r^*_{nm}r_{\nu n}e^{-i(\omega_p-\omega_q-\omega_r+\omega_{\nu g})t}}{(\omega_p+\omega_{mg})(\omega_p-\omega_q+\omega_{ng})((\omega_p-\omega_q-\omega_r+\omega_{\nu g}))}
- \frac{r_{mg}r_{nm}r^*_{\nu n}e^{i(\omega_p+\omega_q-\omega_r-\omega_{\nu g})t}}{(\omega_p-\omega_{mg})(\omega_p+\omega_q-\omega_{ng})((\omega_p+\omega_q-\omega_r-\omega_{\nu g}))}\\
\nonumber
&+&
\frac{r_{mg}r^*_{nm}r^*_{\nu n}e^{i(\omega_p-\omega_q-\omega_r-\omega_{\nu g})t}}{(\omega_p-\omega_{mg})(\omega_p-\omega_q-\omega_{ng})((\omega_p-\omega_q-\omega_r-\omega_{\nu g}))}
+
\frac{r^*_{mg}r_{nm}r^*_{\nu n}e^{-i(\omega_p+\omega_q+\omega_r+\omega_{\nu g})t}}{(\omega_p+\omega_{mg})(\omega_p+\omega_q+\omega_{ng})((\omega_p+\omega_q+\omega_r+\omega_{\nu g}))}\\
&-&\frac{r^*_{mg}r^*_{nm}r^*_{\nu n}e^{-i(\omega_p-\omega_q+\omega_r+\omega_{\nu g})t}}{(\omega_p+\omega_{mg})(\omega_p-\omega_q+\omega_{ng})((\omega_p-\omega_q+\omega_r+\omega_{\nu g}))}\bigg]
\end{eqnarray}
\end{widetext}
The third order optical gyration vector can be written as
\begin{widetext}
\begin{eqnarray}
\nonumber
{\bf m}^{(3)}&=&\frac{Ne}{2m}\big(\frac{e}{2\hbar}\big)^6\sum_{mn\nu}\langle u_{\nu}|(\hat{{\bf L}}+g_s\hat{{\bf S}})|u_{\nu}\rangle|r_{mg}r_{nm}r_{\nu n}|^2\\
\nonumber
&&\Big[
D^{-1}(\omega_p-\omega_{mg})D^{-1}(\omega_p+\omega_q-\omega_{ng})D^{-1}(\omega_p+\omega_q+\omega_r-\omega_{\nu g})\\
\nonumber
&+&
D^{-1}(\omega_p-\omega_{mg})D^{-1}(\omega_p-\omega_q-\omega_{ng})D^{-1}(\omega_p-\omega_q+\omega_r-\omega_{\nu g})\\
\nonumber
&+&
D^{-1}(\omega_p+\omega_{mg})D^{-1}(\omega_p+\omega_q+\omega_{ng})D^{-1}(\omega_p+\omega_q-\omega_r+\omega_{\nu g})\\
\nonumber
&+&
D^{-1}(\omega_p+\omega_{mg})D^{-1}(\omega_p-\omega_q+\omega_{ng})D^{-1}(\omega_p-\omega_q-\omega_r+\omega_{\nu g})\\
\nonumber
&+&
D^{-1}(\omega_p-\omega_{mg})D^{-1}(\omega_p+\omega_q-\omega_{ng})D^{-1}(\omega_p+\omega_q-\omega_r-\omega_{\nu g})\\
\nonumber
&+&
D^{-1}(\omega_p-\omega_{mg})D^{-1}(\omega_p-\omega_q-\omega_{ng})D^{-1}(\omega_p-\omega_q-\omega_r-\omega_{\nu g})\\
\nonumber
&+&
D^{-1}(\omega_p+\omega_{mg})D^{-1}(\omega_p+\omega_q+\omega_{ng})D^{-1}(\omega_p+\omega_q+\omega_r+\omega_{\nu g})\\
&+&
D^{-1}(\omega_p+\omega_{mg})D^{-1}(\omega_p-\omega_q+\omega_{ng})D^{-1}(\omega_p-\omega_q+\omega_r+\omega_{\nu g})\Big]
\end{eqnarray}
\end{widetext}

\bibliography{hamed_biblio}

\begin{thebibliography}{27}%
\makeatletter
\providecommand \@ifxundefined [1]{%
 \@ifx{#1\undefined}
}%
\providecommand \@ifnum [1]{%
 \ifnum #1\expandafter \@firstoftwo
 \else \expandafter \@secondoftwo
 \fi
}%
\providecommand \@ifx [1]{%
 \ifx #1\expandafter \@firstoftwo
 \else \expandafter \@secondoftwo
 \fi
}%
\providecommand \natexlab [1]{#1}%
\providecommand \enquote  [1]{``#1''}%
\providecommand \bibnamefont  [1]{#1}%
\providecommand \bibfnamefont [1]{#1}%
\providecommand \citenamefont [1]{#1}%
\providecommand \href@noop [0]{\@secondoftwo}%
\providecommand \href [0]{\begingroup \@sanitize@url \@href}%
\providecommand \@href[1]{\@@startlink{#1}\@@href}%
\providecommand \@@href[1]{\endgroup#1\@@endlink}%
\providecommand \@sanitize@url [0]{\catcode `\\12\catcode `\$12\catcode
  `\&12\catcode `\#12\catcode `\^12\catcode `\_12\catcode `\%12\relax}%
\providecommand \@@startlink[1]{}%
\providecommand \@@endlink[0]{}%
\providecommand \url  [0]{\begingroup\@sanitize@url \@url }%
\providecommand \@url [1]{\endgroup\@href {#1}{\urlprefix }}%
\providecommand \urlprefix  [0]{URL }%
\providecommand \Eprint [0]{\href }%
\providecommand \doibase [0]{http://dx.doi.org/}%
\providecommand \selectlanguage [0]{\@gobble}%
\providecommand \bibinfo  [0]{\@secondoftwo}%
\providecommand \bibfield  [0]{\@secondoftwo}%
\providecommand \translation [1]{[#1]}%
\providecommand \BibitemOpen [0]{}%
\providecommand \bibitemStop [0]{}%
\providecommand \bibitemNoStop [0]{.\EOS\space}%
\providecommand \EOS [0]{\spacefactor3000\relax}%
\providecommand \BibitemShut  [1]{\csname bibitem#1\endcsname}%
\let\auto@bib@innerbib\@empty
\bibitem [{\citenamefont {Kimel}\ \emph {et~al.}(2004)\citenamefont {Kimel},
  \citenamefont {Kirilyuk}, \citenamefont {Tsvetkov}, \citenamefont {Pisarev},\
  and\ \citenamefont {Rasing}}]{kimel2004laser}%
  \BibitemOpen
  \bibfield  {author} {\bibinfo {author} {\bibfnamefont {A.}~\bibnamefont
  {Kimel}}, \bibinfo {author} {\bibfnamefont {A.}~\bibnamefont {Kirilyuk}},
  \bibinfo {author} {\bibfnamefont {A.}~\bibnamefont {Tsvetkov}}, \bibinfo
  {author} {\bibfnamefont {R.}~\bibnamefont {Pisarev}}, \ and\ \bibinfo
  {author} {\bibfnamefont {T.}~\bibnamefont {Rasing}},\ }\href@noop {}
  {\bibfield  {journal} {\bibinfo  {journal} {Nature}\ }\textbf {\bibinfo
  {volume} {429}},\ \bibinfo {pages} {850} (\bibinfo {year}
  {2004})}\BibitemShut {NoStop}%
\bibitem [{\citenamefont {Kimel}\ \emph {et~al.}(2005)\citenamefont {Kimel},
  \citenamefont {Kirilyuk}, \citenamefont {Usachev}, \citenamefont {Pisarev},
  \citenamefont {Balbashov},\ and\ \citenamefont
  {Rasing}}]{kimel2005ultrafast}%
  \BibitemOpen
  \bibfield  {author} {\bibinfo {author} {\bibfnamefont {A.}~\bibnamefont
  {Kimel}}, \bibinfo {author} {\bibfnamefont {A.}~\bibnamefont {Kirilyuk}},
  \bibinfo {author} {\bibfnamefont {P.}~\bibnamefont {Usachev}}, \bibinfo
  {author} {\bibfnamefont {R.}~\bibnamefont {Pisarev}}, \bibinfo {author}
  {\bibfnamefont {A.}~\bibnamefont {Balbashov}}, \ and\ \bibinfo {author}
  {\bibfnamefont {T.}~\bibnamefont {Rasing}},\ }\href@noop {} {\bibfield
  {journal} {\bibinfo  {journal} {Nature}\ }\textbf {\bibinfo {volume} {435}},\
  \bibinfo {pages} {655} (\bibinfo {year} {2005})}\BibitemShut {NoStop}%
\bibitem [{\citenamefont {Kirilyuk}\ \emph {et~al.}(2010)\citenamefont
  {Kirilyuk}, \citenamefont {Kimel},\ and\ \citenamefont
  {Rasing}}]{kirilyuk2010ultrafast}%
  \BibitemOpen
  \bibfield  {author} {\bibinfo {author} {\bibfnamefont {A.}~\bibnamefont
  {Kirilyuk}}, \bibinfo {author} {\bibfnamefont {A.~V.}\ \bibnamefont {Kimel}},
  \ and\ \bibinfo {author} {\bibfnamefont {T.}~\bibnamefont {Rasing}},\
  }\href@noop {} {\bibfield  {journal} {\bibinfo  {journal} {Reviews of Modern
  Physics}\ }\textbf {\bibinfo {volume} {82}},\ \bibinfo {pages} {2731}
  (\bibinfo {year} {2010})}\BibitemShut {NoStop}%
\bibitem [{\citenamefont {Subkhangulov}\ \emph {et~al.}(2016)\citenamefont
  {Subkhangulov}, \citenamefont {Mikhaylovskiy}, \citenamefont {Zvezdin},
  \citenamefont {Kruglyak}, \citenamefont {Rasing},\ and\ \citenamefont
  {Kimel}}]{subkhangulov2016terahertz}%
  \BibitemOpen
  \bibfield  {author} {\bibinfo {author} {\bibfnamefont {R.}~\bibnamefont
  {Subkhangulov}}, \bibinfo {author} {\bibfnamefont {R.}~\bibnamefont
  {Mikhaylovskiy}}, \bibinfo {author} {\bibfnamefont {A.}~\bibnamefont
  {Zvezdin}}, \bibinfo {author} {\bibfnamefont {V.}~\bibnamefont {Kruglyak}},
  \bibinfo {author} {\bibfnamefont {T.}~\bibnamefont {Rasing}}, \ and\ \bibinfo
  {author} {\bibfnamefont {A.}~\bibnamefont {Kimel}},\ }\href@noop {}
  {\bibfield  {journal} {\bibinfo  {journal} {Nature Photonics}\ }\textbf
  {\bibinfo {volume} {10}},\ \bibinfo {pages} {111} (\bibinfo {year}
  {2016})}\BibitemShut {NoStop}%
\bibitem [{\citenamefont {Ghamsari}\ and\ \citenamefont
  {Berini}(2016)}]{ghamsari2016nonlinear}%
  \BibitemOpen
  \bibfield  {author} {\bibinfo {author} {\bibfnamefont {B.~G.}\ \bibnamefont
  {Ghamsari}}\ and\ \bibinfo {author} {\bibfnamefont {P.}~\bibnamefont
  {Berini}},\ }\href@noop {} {\bibfield  {journal} {\bibinfo  {journal} {Nature
  Photonics}\ }\textbf {\bibinfo {volume} {10}},\ \bibinfo {pages} {74}
  (\bibinfo {year} {2016})}\BibitemShut {NoStop}%
\bibitem [{\citenamefont {Pitaevskii}(1961)}]{pitaevskii1961electric}%
  \BibitemOpen
  \bibfield  {author} {\bibinfo {author} {\bibfnamefont {L.}~\bibnamefont
  {Pitaevskii}},\ }\href@noop {} {\bibfield  {journal} {\bibinfo  {journal}
  {Sov. Phys. JETP}\ }\textbf {\bibinfo {volume} {12}},\ \bibinfo {pages}
  {1008} (\bibinfo {year} {1961})}\BibitemShut {NoStop}%
\bibitem [{\citenamefont {Van~der Ziel}\ \emph {et~al.}(1965)\citenamefont
  {Van~der Ziel}, \citenamefont {Pershan},\ and\ \citenamefont
  {Malmstrom}}]{van1965optically}%
  \BibitemOpen
  \bibfield  {author} {\bibinfo {author} {\bibfnamefont {J.}~\bibnamefont
  {Van~der Ziel}}, \bibinfo {author} {\bibfnamefont {P.~S.}\ \bibnamefont
  {Pershan}}, \ and\ \bibinfo {author} {\bibfnamefont {L.}~\bibnamefont
  {Malmstrom}},\ }\href@noop {} {\bibfield  {journal} {\bibinfo  {journal}
  {Physical review letters}\ }\textbf {\bibinfo {volume} {15}},\ \bibinfo
  {pages} {190} (\bibinfo {year} {1965})}\BibitemShut {NoStop}%
\bibitem [{\citenamefont {Pershan}\ \emph {et~al.}(1966)\citenamefont
  {Pershan}, \citenamefont {Van~der Ziel},\ and\ \citenamefont
  {Malmstrom}}]{pershan1966theoretical}%
  \BibitemOpen
  \bibfield  {author} {\bibinfo {author} {\bibfnamefont {P.}~\bibnamefont
  {Pershan}}, \bibinfo {author} {\bibfnamefont {J.}~\bibnamefont {Van~der
  Ziel}}, \ and\ \bibinfo {author} {\bibfnamefont {L.}~\bibnamefont
  {Malmstrom}},\ }\href@noop {} {\bibfield  {journal} {\bibinfo  {journal}
  {Physical review}\ }\textbf {\bibinfo {volume} {143}},\ \bibinfo {pages}
  {574} (\bibinfo {year} {1966})}\BibitemShut {NoStop}%
\bibitem [{\citenamefont {Kalashnikova}\ \emph {et~al.}(2015)\citenamefont
  {Kalashnikova}, \citenamefont {Kimel},\ and\ \citenamefont
  {Pisarev}}]{kalashnikova2015ultrafast}%
  \BibitemOpen
  \bibfield  {author} {\bibinfo {author} {\bibfnamefont {A.~M.}\ \bibnamefont
  {Kalashnikova}}, \bibinfo {author} {\bibfnamefont {A.~V.}\ \bibnamefont
  {Kimel}}, \ and\ \bibinfo {author} {\bibfnamefont {R.~V.}\ \bibnamefont
  {Pisarev}},\ }\href@noop {} {\bibfield  {journal} {\bibinfo  {journal}
  {Physics-Uspekhi}\ }\textbf {\bibinfo {volume} {58}},\ \bibinfo {pages} {969}
  (\bibinfo {year} {2015})}\BibitemShut {NoStop}%
\bibitem [{\citenamefont {Popova}\ \emph {et~al.}(2011)\citenamefont {Popova},
  \citenamefont {Bringer},\ and\ \citenamefont
  {Bl{\"u}gel}}]{popova2011theory}%
  \BibitemOpen
  \bibfield  {author} {\bibinfo {author} {\bibfnamefont {D.}~\bibnamefont
  {Popova}}, \bibinfo {author} {\bibfnamefont {A.}~\bibnamefont {Bringer}}, \
  and\ \bibinfo {author} {\bibfnamefont {S.}~\bibnamefont {Bl{\"u}gel}},\
  }\href@noop {} {\bibfield  {journal} {\bibinfo  {journal} {Physical Review
  B}\ }\textbf {\bibinfo {volume} {84}},\ \bibinfo {pages} {214421} (\bibinfo
  {year} {2011})}\BibitemShut {NoStop}%
\bibitem [{\citenamefont {Taguchi}\ and\ \citenamefont
  {Tatara}(2011)}]{taguchi2011theory}%
  \BibitemOpen
  \bibfield  {author} {\bibinfo {author} {\bibfnamefont {K.}~\bibnamefont
  {Taguchi}}\ and\ \bibinfo {author} {\bibfnamefont {G.}~\bibnamefont
  {Tatara}},\ }\href@noop {} {\bibfield  {journal} {\bibinfo  {journal}
  {Physical Review B}\ }\textbf {\bibinfo {volume} {84}},\ \bibinfo {pages}
  {174433} (\bibinfo {year} {2011})}\BibitemShut {NoStop}%
\bibitem [{\citenamefont {Taguchi}\ and\ \citenamefont
  {Tatara}(2012)}]{taguchi2012theory}%
  \BibitemOpen
  \bibfield  {author} {\bibinfo {author} {\bibfnamefont {K.}~\bibnamefont
  {Taguchi}}\ and\ \bibinfo {author} {\bibfnamefont {G.}~\bibnamefont
  {Tatara}},\ }in\ \href@noop {} {\emph {\bibinfo {booktitle} {Journal of
  Physics: Conference Series}}},\ Vol.\ \bibinfo {volume} {400}\ (\bibinfo
  {organization} {IOP Publishing},\ \bibinfo {year} {2012})\ p.\ \bibinfo
  {pages} {042055}\BibitemShut {NoStop}%
\bibitem [{\citenamefont {Battiato}\ \emph {et~al.}(2012)\citenamefont
  {Battiato}, \citenamefont {Barbalinardo}, \citenamefont {Carva},\ and\
  \citenamefont {Oppeneer}}]{battiato2012beyond}%
  \BibitemOpen
  \bibfield  {author} {\bibinfo {author} {\bibfnamefont {M.}~\bibnamefont
  {Battiato}}, \bibinfo {author} {\bibfnamefont {G.}~\bibnamefont
  {Barbalinardo}}, \bibinfo {author} {\bibfnamefont {K.}~\bibnamefont {Carva}},
  \ and\ \bibinfo {author} {\bibfnamefont {P.~M.}\ \bibnamefont {Oppeneer}},\
  }\href@noop {} {\bibfield  {journal} {\bibinfo  {journal} {Physical Review
  B}\ }\textbf {\bibinfo {volume} {85}},\ \bibinfo {pages} {045117} (\bibinfo
  {year} {2012})}\BibitemShut {NoStop}%
\bibitem [{\citenamefont {Battiato}\ \emph {et~al.}(2014)\citenamefont
  {Battiato}, \citenamefont {Barbalinardo},\ and\ \citenamefont
  {Oppeneer}}]{battiato2014quantum}%
  \BibitemOpen
  \bibfield  {author} {\bibinfo {author} {\bibfnamefont {M.}~\bibnamefont
  {Battiato}}, \bibinfo {author} {\bibfnamefont {G.}~\bibnamefont
  {Barbalinardo}}, \ and\ \bibinfo {author} {\bibfnamefont {P.~M.}\
  \bibnamefont {Oppeneer}},\ }\href@noop {} {\bibfield  {journal} {\bibinfo
  {journal} {Physical review B}\ }\textbf {\bibinfo {volume} {89}},\ \bibinfo
  {pages} {014413} (\bibinfo {year} {2014})}\BibitemShut {NoStop}%
\bibitem [{\citenamefont {Boyd}(2019)}]{boyd2019nonlinear}%
  \BibitemOpen
  \bibfield  {author} {\bibinfo {author} {\bibfnamefont {R.~W.}\ \bibnamefont
  {Boyd}},\ }\href@noop {} {\emph {\bibinfo {title} {Nonlinear optics}}}\
  (\bibinfo  {publisher} {Academic press},\ \bibinfo {year} {2019})\
  Chap.~\bibinfo {chapter} {3}\BibitemShut {NoStop}%
\bibitem [{\citenamefont {Band}(2006)}]{band2006light}%
  \BibitemOpen
  \bibfield  {author} {\bibinfo {author} {\bibfnamefont {Y.~B.}\ \bibnamefont
  {Band}},\ }\href@noop {} {\emph {\bibinfo {title} {Light and matter:
  electromagnetism, optics, spectroscopy and lasers}}},\ Vol.~\bibinfo {volume}
  {1}\ (\bibinfo {year} {2006})\BibitemShut {NoStop}%
\bibitem [{\citenamefont {Zon}\ and\ \citenamefont
  {Kupershmidt}(1976)}]{zon1976inverse}%
  \BibitemOpen
  \bibfield  {author} {\bibinfo {author} {\bibfnamefont {B.}~\bibnamefont
  {Zon}}\ and\ \bibinfo {author} {\bibfnamefont {V.~Y.}\ \bibnamefont
  {Kupershmidt}},\ }\href@noop {} {\bibfield  {journal} {\bibinfo  {journal}
  {Radiophysics and Quantum Electronics}\ }\textbf {\bibinfo {volume} {19}},\
  \bibinfo {pages} {1072} (\bibinfo {year} {1976})}\BibitemShut {NoStop}%
\bibitem [{\citenamefont {Hertel}(2006)}]{hertel2006theory}%
  \BibitemOpen
  \bibfield  {author} {\bibinfo {author} {\bibfnamefont {R.}~\bibnamefont
  {Hertel}},\ }\href@noop {} {\bibfield  {journal} {\bibinfo  {journal}
  {Journal of magnetism and magnetic materials}\ }\textbf {\bibinfo {volume}
  {303}},\ \bibinfo {pages} {L1} (\bibinfo {year} {2006})}\BibitemShut
  {NoStop}%
\bibitem [{\citenamefont {Hertel}\ and\ \citenamefont
  {F{\"a}hnle}(2015)}]{hertel2015macroscopic}%
  \BibitemOpen
  \bibfield  {author} {\bibinfo {author} {\bibfnamefont {R.}~\bibnamefont
  {Hertel}}\ and\ \bibinfo {author} {\bibfnamefont {M.}~\bibnamefont
  {F{\"a}hnle}},\ }\href@noop {} {\bibfield  {journal} {\bibinfo  {journal}
  {Physical Review B}\ }\textbf {\bibinfo {volume} {91}},\ \bibinfo {pages}
  {020411} (\bibinfo {year} {2015})}\BibitemShut {NoStop}%
\bibitem [{\citenamefont {Bloembergen}(1996)}]{bloembergen1996encounters}%
  \BibitemOpen
  \bibfield  {author} {\bibinfo {author} {\bibfnamefont {N.}~\bibnamefont
  {Bloembergen}},\ }\href@noop {} {\emph {\bibinfo {title} {Encounters In
  Nonlinear Optics: Selected Papers of Nicolaas Bloembergen (With
  Commentary)}}}\ (\bibinfo  {publisher} {World Scientific},\ \bibinfo {year}
  {1996})\ Chap.~\bibinfo {chapter} {1}\BibitemShut {NoStop}%
\bibitem [{\citenamefont {Jackson}(1999)}]{jackson1999classical}%
  \BibitemOpen
  \bibfield  {author} {\bibinfo {author} {\bibfnamefont {J.~D.}\ \bibnamefont
  {Jackson}},\ }\href@noop {} {\emph {\bibinfo {title} {Classical
  electrodynamics}}}\ (\bibinfo  {publisher} {American Association of Physics
  Teachers},\ \bibinfo {year} {1999})\ Chap.~\bibinfo {chapter} {5}\BibitemShut
  {NoStop}%
\bibitem [{\citenamefont {Landau}\ \emph {et~al.}(2013)\citenamefont {Landau},
  \citenamefont {Bell}, \citenamefont {Kearsley}, \citenamefont {Pitaevskii},
  \citenamefont {Lifshitz},\ and\ \citenamefont
  {Sykes}}]{landau2013electrodynamics}%
  \BibitemOpen
  \bibfield  {author} {\bibinfo {author} {\bibfnamefont {L.~D.}\ \bibnamefont
  {Landau}}, \bibinfo {author} {\bibfnamefont {J.}~\bibnamefont {Bell}},
  \bibinfo {author} {\bibfnamefont {M.}~\bibnamefont {Kearsley}}, \bibinfo
  {author} {\bibfnamefont {L.}~\bibnamefont {Pitaevskii}}, \bibinfo {author}
  {\bibfnamefont {E.}~\bibnamefont {Lifshitz}}, \ and\ \bibinfo {author}
  {\bibfnamefont {J.}~\bibnamefont {Sykes}},\ }\href@noop {} {\emph {\bibinfo
  {title} {Electrodynamics of continuous media}}},\ Vol.~\bibinfo {volume} {8}\
  (\bibinfo  {publisher} {elsevier},\ \bibinfo {year} {2013})\ Chap.~\bibinfo
  {chapter} {XI}\BibitemShut {NoStop}%
\bibitem [{\citenamefont {Saleh}\ and\ \citenamefont
  {Teich}(2019)}]{saleh2019fundamentals}%
  \BibitemOpen
  \bibfield  {author} {\bibinfo {author} {\bibfnamefont {B.~E.}\ \bibnamefont
  {Saleh}}\ and\ \bibinfo {author} {\bibfnamefont {M.~C.}\ \bibnamefont
  {Teich}},\ }\href@noop {} {\emph {\bibinfo {title} {Fundamentals of
  photonics}}}\ (\bibinfo  {publisher} {john Wiley \& sons},\ \bibinfo {year}
  {2019})\ Chap.~\bibinfo {chapter} {6}\BibitemShut {NoStop}%
\bibitem [{\citenamefont {Im}\ \emph {et~al.}(2017)\citenamefont {Im},
  \citenamefont {Ri}, \citenamefont {Ho},\ and\ \citenamefont
  {Herrmann}}]{im2017third}%
  \BibitemOpen
  \bibfield  {author} {\bibinfo {author} {\bibfnamefont {S.-J.}\ \bibnamefont
  {Im}}, \bibinfo {author} {\bibfnamefont {C.-S.}\ \bibnamefont {Ri}}, \bibinfo
  {author} {\bibfnamefont {K.-S.}\ \bibnamefont {Ho}}, \ and\ \bibinfo {author}
  {\bibfnamefont {J.}~\bibnamefont {Herrmann}},\ }\href@noop {} {\bibfield
  {journal} {\bibinfo  {journal} {Physical Review B}\ }\textbf {\bibinfo
  {volume} {96}},\ \bibinfo {pages} {165437} (\bibinfo {year}
  {2017})}\BibitemShut {NoStop}%
\bibitem [{\citenamefont {Shen}(1984)}]{shen1984principles}%
  \BibitemOpen
  \bibfield  {author} {\bibinfo {author} {\bibfnamefont {Y.-R.}\ \bibnamefont
  {Shen}},\ }\href@noop {} {\emph {\bibinfo {title} {The principles of
  nonlinear optics}}}\ (\bibinfo {year} {1984})\ Chap.\ \bibinfo {chapter}
  {4,5}\BibitemShut {NoStop}%
\bibitem [{\citenamefont {Voigt}(1902)}]{voigt1902vii}%
  \BibitemOpen
  \bibfield  {author} {\bibinfo {author} {\bibfnamefont {W.}~\bibnamefont
  {Voigt}},\ }\href@noop {} {\bibfield  {journal} {\bibinfo  {journal} {The
  London, Edinburgh, and Dublin Philosophical Magazine and Journal of Science}\
  }\textbf {\bibinfo {volume} {4}},\ \bibinfo {pages} {90} (\bibinfo {year}
  {1902})}\BibitemShut {NoStop}%
\bibitem [{\citenamefont {Cotton}\ and\ \citenamefont
  {Mouton}(1907)}]{cotton1907new}%
  \BibitemOpen
  \bibfield  {author} {\bibinfo {author} {\bibfnamefont {A.}~\bibnamefont
  {Cotton}}\ and\ \bibinfo {author} {\bibfnamefont {H.}~\bibnamefont
  {Mouton}},\ }\href@noop {} {\bibfield  {journal} {\bibinfo  {journal} {CR
  Hebd Seances Acad. Sci. Paris}\ }\textbf {\bibinfo {volume} {145}},\ \bibinfo
  {pages} {229} (\bibinfo {year} {1907})}\BibitemShut {NoStop}%
\end{thebibliography}%

\end{document}